\documentclass[aps,prl,tightenlines,onecolumn,superscriptaddress,showpacs]{revtex4}
\usepackage{amssymb}
\usepackage{color,graphicx}
\usepackage{lmodern}
\usepackage{amsmath}
\usepackage{amsbsy}
\usepackage{amsthm}
\usepackage{float}
\usepackage{bbm}
\usepackage{bm}
\usepackage{subfigure}
\usepackage[paperwidth=210mm,paperheight=297mm,centering,hmargin=1.5cm,vmargin=2cm]{geometry}

\begin{document}
\title{Perturbative Treatment of Quantum to Classical Transition in Chiral Molecules:\\ Dilute Phase vs. Condensed Phase}

\author{Farhad Taher Ghahramani}
\email[Corresponding Author:~]{farhadtqm@ipm.ir}
\affiliation{Foundations of Physics Group, School of Physics, Institute for Research in Fundamental Sciences (IPM), P.O. Box 19395-5531, Tehran, Iran}
\author{Arash Tirandaz}
\email[]{arash85201@ipm.ir}
\affiliation{Foundations of Physics Group, School of Physics, Institute for Research in Fundamental Sciences (IPM), P.O. Box 19395-5531, Tehran, Iran}

\begin{abstract}
  We examine the dynamics of chiral states of chiral molecules with high tunneling rates in dilute and condensed phases in the context of time-dependent perturbation theory. The chiral molecule is effectively described by an asymmetric double-well potential, whose asymmetry is a measure of chiral interactions. The dilute and condensed phases are conjointly described by a collection of harmonic oscillators but respectively with temperature-dependent sub-ohmic and temperature-independent ohmic spectral densities. We examine our method quantitatively by applying the dynamics to isotopic ammonia molecule, NHDT, in an inert background gas (as the dilute phase) and in water (as the condensed phase). As different spectral densities implies, the extension of the dynamics from the dilute phase to the condensed phase is not trivial. While the dynamics in the dilute phase leads to racemization, the chiral interactions in the condensed phase induce the quantum Zeno effect. Moreover, contrary to the condensed phase, the short-time dynamics in the dilute phase is sensitive to the initial state of the chiral molecule and to the strength of the coupling between the molecule and the environment.
\end{abstract}
\maketitle
\section{I. Introduction}
One of the fundamental problems in molecular science is the quantum-mechanical origin of chirality. The chiral configurations of a chiral molecule have a many-particle, complex dynamics. This dynamics can be effectively modeled by the motion of a particle in a symmetric double-well potential~\cite{Hund}. The states associated to chiral configurations are assumed to be localized in two minima of the potential. The superposition of chiral states is realized by tunneling through the barrier. If the barrier is high enough to prevent tunneling, once prepared, the chiral state is preserved. Upon a quantitative analysis, however, this explanation seems rather insufficient for low-barrier molecules~\cite{Jan}. As a modern resolution, the chiral interactions, i.e. interactions that prefer a particular chiral state, became important. They can be effectively taken into account by introducing an asymmetry parameter into the symmetric double-well potential. If they are strong enough to overcome tunneling process, the molecule is confined in the preferred chiral state. The most discussed interactions in this context are parity-violating weak interactions~\cite{Let,Has}, intermolecular interactions~\cite{Sto} and interaction with the circularly-polarized light~\cite{Nor}. The problem is that the induced chirality is demolished by the dissipative effects of the surrounding environment. The open chiral molecule has been studied by the mean-field approach, and more completely by the decoherence theory. In the mean-field theory, the environment is envisaged as an effective potential added to the Schr\"{o}dinger dynamics of the system~\cite{Bre}. The resulting non-linear dynamics can be served to stabilize a particular chiral state~\cite{Var,Jon,Gre}. Recently, this approach is extended to include chiral interactions in the Langevin formalism of open quantum systems~\cite{Lan1,Lan2,Lan3,Lan4,Lan5}. \\
\indent The decoherence theory describe the superselection of a preferred set of system's states by the environment-induced, dynamical destruction of quantum coherence between them~\cite{Bre,Sch}. The environment acting upon the molecule might be dilute (e.g., gaseous environment) or condensed (e.g., biological environment). The dynamics of a system in a gas phase was studied by the scattering model or more conveniently collisional decoherence~\cite{Joos,Dio,Adl,Vac}. The application of collisional decoherence to the study of the tunneling of a chiral molecule in a gas phase was first done in the pioneering works of Harris and Stodolsky~\cite{Har1,Har2} and Silbey and Harris~\cite{Sil}. More refined works showed that the environmental collisions induce an indirect position-measurement on the molecule and thus stabilize the chiral states~\cite{Hor,Bah1,Bah2,Gon,Gha}. In a condensed phase, however, since the medium is always present, the idea of a collision is inapplicable. The simplest representation of a condensed phase is a bath of harmonic oscillators. At sufficiently low temperatures, the states of the molecule are effectively confined in the two-dimensional Hilbert space spanned by the two chiral states. When the two-level system couples linearly to the environmental oscillators, the result is the renowned Spin-Boson model studied extensively in the literature, especially by Leggett and co-workers\cite{Leg}. Recently, this model is employed to examine the role of quantum coherence in biological systems~\cite{Gil,Raz}. An elementary application of this model to chiral molecules is found in the work of Silbey and Harris~\cite{Sil}. Recently, Tirandaz and co-workers solved the general Spin-Boson model to examine the interplay between tunneling process and chiral interactions in a biological environment~\cite{Tir}. They showed that despite the powerful racemization effects of the biological environment the chiral interactions induced by the same environment can prohibit the tunneling process and thus preserve the initial chiral state.\\
\indent The description of the decoherence dynamics is usually formulated in terms of the so-called master equations, which yield the {\it approximate} evolution of the reduced density matrix of the open quantum system. The master equation is essentially a mathematical object which maps the initial state to the final state without providing a clear-cut physical understanding of the inner dynamics. Moreover, it is recently discussed that conceptually the state of the closed system and the reduced state assigned to that system when it interacts with the environment are different~\cite{For1,For2}. In order to present a simple and straightforward demonstration of the stabilization of molecular chirality, in this paper, we apply the time-dependent perturbation theory to examine the dynamics of an open chiral molecule. The main emphasis here would be on the comparison between dynamics influenced by dilute and condensed environments. It is demonstrated that the interaction with {\it any} environment can be rigorously mapped onto a system linearly coupled to an oscillator environment, provided the interaction is sufficiently weak and second-order perturbation theory can be applied~\cite{Fey,Cal}. We describe the dynamics of chirality by an asymmetric double-well potential, and the environment is represented by a collection of harmonic oscillators, with different spectral densities for dilute and condensed phases. We exemplify the dilute and condensed phases respectively by an inert background gas and water. Note that for we focus on chiral molecules with high tunneling rates, the strength of chiral interactions can be considered less than the tunneling strength, especially in the dilute phase.\\
\indent The paper is organized as follows. In the next section, we describe the chiral molecule and its harmonic environment. We examine the dynamics of chiral states in the third section using time-dependent perturbation theory. In the forth section, we first estimate the parameters relevant to our analysis, specify the spectral densities of dilute and condensed environments and accordingly present our results. Finally, we summarize our findings in the last section.
\section{II. Model}
The total Hamiltonian of the entire system composed of the chiral molecule and the environmental particles is conveniently defined as
\begin{equation}\label{1}
H=H_{M}+H_{E}+H_{ME}
\end{equation}
where $H_{M}$, $H_{E}$ and $H_{ME}$ are the molecular, environmental and interaction Hamiltonians, respectively.\\
\indent A chiral molecule is found at least as two identical enantiomers through the inversion at molecule's center of mass by a long-amplitude vibration known as the contortional vibration~\cite{Tow,Her}. This particular vibration can effectively be described by the motion of a particle in an asymmetric double-well potential. The minima of the potential are associated to two enantiomers of the molecule and the asymmetry can be considered as an overall measure of all chiral interactions. For an isolated chiral molecule, the asymmetry is merely resulted from the internal chiral interactions i.e., the parity-violating weak interactions~\cite{Let,Qua}. For a chiral molecule in interaction with the environment, however, the external chiral interactions e.g., the dispersion intermolecular interactions~\cite{Sto} and interaction with the circularly-polarized light~\cite{Gha} are the main source of asymmetry. An asymmetric double-well potential can be represented by a quartic potential including a linear term
\begin{equation}\label{2}
U(R)=U_{\mbox{\tiny${\mbox{\tiny$\circ$}}$}}\Big\{\Big[\big(\frac{R}{R_{\mbox{\tiny${\mbox{\tiny$\circ$}}$}}}\big)^{2}-1\Big]^{2}-1-\eta\big(\frac{R}{R_{\mbox{\tiny${\mbox{\tiny$\circ$}}$}}}\big)\Big\},\qquad U_{\mbox{\tiny${\mbox{\tiny$\circ$}}$}}=\frac{M\Omega^{2}R_{\mbox{\tiny${\mbox{\tiny$\circ$}}$}}^{2}}{8}
\end{equation}
where $M$ is the effective mass of the chiral molecule, $\Omega$ is the harmonic frequency at the bottom of each well, $\eta$ is called the asymmetry parameter and $R_{\mbox{\tiny${\mbox{\tiny$\circ$}}$}}$ is the distance between two minima from the origin for $\eta=0$. To quantify the extent to which the molecule exhibits quantum coherence, we incorporate the dimensionless form of the model. The potential has the characteristic length $R_{\mbox{\tiny${\mbox{\tiny$\circ$}}$}}$ and the characteristic energy $U_{\mbox{\tiny${\mbox{\tiny$\circ$}}$}}$ which we adopt as the units of length and energy, respectively. The corresponding characteristic time can be defined as $\tau_{\mbox{\tiny${\mbox{\tiny$\circ$}}$}}=R_{\mbox{\tiny${\mbox{\tiny$\circ$}}$}}/(U_{\mbox{\tiny${\mbox{\tiny$\circ$}}$}}/M)^{1/2}$ which we consider as the unit of the time. Likewise, the unit of the momentum is taken as $P_{\mbox{\tiny${\mbox{\tiny$\circ$}}$}}=(MU_{\mbox{\tiny${\mbox{\tiny$\circ$}}$}})^{1/2}$. We then define the dynamical variables, $x$ and $p$, as $R/R_{\mbox{\tiny${\mbox{\tiny$\circ$}}$}}$ and $P/P_{\mbox{\tiny${\mbox{\tiny$\circ$}}$}}$, respectively. The corresponding commutation relation is defined as $[x,p]=\imath h$, where Planck constant is redefined as $h=\hbar/R_{\mbox{\tiny${\mbox{\tiny$\circ$}}$}}P_{\mbox{\tiny${\mbox{\tiny$\circ$}}$}}=\hbar/U_{\mbox{\tiny${\mbox{\tiny$\circ$}}$}}\tau_{\mbox{\tiny${\mbox{\tiny$\circ$}}$}}$, which we call the reduced Planck constant. \\
\indent In the limit $E_{th}\ll\Omega \tau_{\mbox{\tiny${\mbox{\tiny$\circ$}}$}}h\ll 1$ (with $E_{th}=k_{\mbox{\tiny$B$}}T/U_{\mbox{\tiny${\mbox{\tiny$\circ$}}$}}$ as thermal energy, $k_{\mbox{\tiny$B$}}$ as Boltzmann constant, and $T$ as temperature), energy states are confined in two-dimensional Hilbert space spanned by two chiral states $|L\rangle$ and $|R\rangle$. In fact, such a two-level approximation holds for most chiral molecules~\cite{Tow,Her}. Accordingly, the effective molecular Hamiltonian in the chiral basis can be written as~\cite{Leg}
\begin{equation}\label{3}
H_{M}=-\Delta\sigma_{x}-\delta\sigma_{z}
\end{equation}
where we defined
\begin{align}\label{4}
   \Delta&=\frac{h\Omega\tau_{\mbox{\tiny${\mbox{\tiny$\circ$}}$}}}{4}  \nonumber \\
   \delta&=\eta\sqrt{\frac{h}{2\Omega \tau_{\mbox{\tiny${\mbox{\tiny$\circ$}}$}}}}
\end{align}
in which $\Delta$ is the tunneling strength and $\delta$, known as the localization strength, is a measure of the the energy difference between two enantiomers. If chiral interactions are strong enough to overcome tunneling process, $\Delta\ll\delta$, the chiral states become stable, and thus the chirality problem is resolved. Therefor, here we focus on chiral molecules with low barrier, where the tunneling strength is larger than the localization strength, $\Delta\gg\delta$. The states of energy are described by the superposition of localized states as
\begin{align}\label{5}
|1\rangle&=\cos{\theta}|R\rangle+\sin{\theta}|L\rangle \nonumber \\
|2\rangle&=\sin{\theta}|R\rangle-\cos{\theta}|L\rangle
\end{align}
where we defined $\theta=\frac{1}{2}\arctan\{\Delta/\delta\}$. The corresponding energies are $\mp\frac{1}{2}(\Delta^{2}+\delta^{2})^{1/2}$. For an isolated chiral molecule, the probability of the tunnelling from the left state to the right one is given by
\begin{equation}\label{6}
P_{L\rightarrow R}=\frac{\Delta^{2}}{\Delta^{2}+\delta^{2}}\sin^{2}\{(\Delta^{2}+\delta^{2})^{1/2}t/2\}
\end{equation}
\indent A frequently employed model for an environment is a set of harmonic oscillators. The $\alpha$-th harmonic oscillator in the environment is characterized by its natural frequency, $\omega_{\alpha}$, and position and momentum operators, $x_{\alpha}$ and, $p_{\alpha}$, respectively, according to the Hamiltonian
\begin{equation}\label{7}
H_{E}=\sum_{\alpha}\Big(\frac{1}{2}p_{\alpha}^{2}+\frac{1}{2}\omega_{\alpha}^{2}x_{\alpha}^{2}-\frac{1}{2}h\omega_{\alpha}\Big)
\end{equation}
The last term, which merely displaces the origin of energy, is introduced for later convenience. We define $|vac\rangle$ as the vacuum eigenstate and $|\alpha\rangle$ as an excited eigenstate of the environmental Hamiltonian with energy $\varepsilon_{\alpha}$.\\
\indent The interaction Hamiltonian is generally defined by a linearly-coupled harmonic-environment model as
\begin{equation}\label{8}
H_{ME}=-\sum_{\alpha}\Big(\omega_{\alpha}^{2}f_{\alpha}(\sigma_{z})x_{\alpha}+\frac{1}{2}\omega_{\alpha}^{2}f_{\alpha}^{2}(\sigma_{z})\Big)
\end{equation}
where the function $f_{\alpha}(\sigma_{z})$ is the displacement in each environmental oscillator $\alpha$ induced by the molecule. For simplicity, we assume that the interaction model is separable ($f_{\alpha}(\sigma_{z})=\gamma_{\alpha}f(\sigma_{z})$, where $\gamma_{\alpha}$ is the coupling strength) and also bilinear ($f(\sigma_{z})=\sigma_{z}$).\\
\indent The shift in the molecular energy due to the perturbation $H_{ME}$ is obtained up to the second order as
\begin{align}\label{9}
\delta E_{n}&\simeq\langle n,vac|H_{ME}|n,vac\rangle+\sum_{\substack{m\neq n\\ \alpha\neq vac}}\frac{|\langle m,\alpha|H_{ME}|n,vac\rangle|^{2}}{E_{n}-(E_{m}+\varepsilon_{\alpha})}\nonumber\\
&=\frac{1}{2}\sum_{r}|\sigma_{rn}|^{2}\Omega_{rn}\sum_{\alpha}\frac{\gamma_{\alpha}^{2}\omega_{\alpha}^{3}}{\omega_{\alpha}(\omega_{\alpha}+\Omega_{rn})}
\end{align}
where $\sigma_{mn}=\langle m|\sigma_{z}|n\rangle$, $\Omega_{mn}=E_{m}-E_{n}/h$. The state with the energy shifted to $E_{n}+\delta E_{n}$ by perturbation is not actually stationary and decays with a finite lifetime $\Gamma_{n}^{-1}$ given by Fermi's golden rule as
\begin{align}\label{10}
  \Gamma_{n}&\simeq\frac{2\pi}{h}\sum_{\substack{m\neq n\\ \alpha\neq vac}}|\langle n,vac|H_{ME}|n,vac\rangle|^{2}\delta(E_{n}-(E_{m}+\varepsilon_{\alpha}))  \nonumber\\
  &=\frac{\pi}{h}\sum_{r}|\sigma_{rn}|^{2}\Omega_{rn}\sum_{\alpha}
  \frac{\gamma_{\alpha}^{2}\omega_{\alpha}^{3}\delta(\Omega_{rn}-\omega_{\alpha})}{\omega_{\alpha}}
\end{align}
\section{III. Dynamics}
We assume that the initial state of the entire system can be written as
\begin{equation}\label{11}
|\Psi(0)\rangle=|\psi\rangle|vac\rangle
\end{equation}
where $|\psi\rangle$ is an arbitrary state of the chiral molecule. The state of the total system at time $t$ is obtained by
\begin{equation}\label{12}
|\Psi(t)\rangle=e^{-\imath H_{\mbox{\tiny${\mbox{\tiny$\circ$}}$}}t/h}U_{I}(t)|\Psi(0)\rangle
\end{equation}
where we defined $H_{\mbox{\tiny${\mbox{\tiny$\circ$}}$}}=H_{M}+H_{E}$ and $U_{I}$ is the time evolution operator in the interaction picture: $U_{I}(t)=e^{\imath H_{\mbox{\tiny${\mbox{\tiny$\circ$}}$}}t/h}e^{-\imath Ht/h}$. We expand $U_{I}(t)$ up to the second order with respect to the interaction Hamiltonian to find
\begin{equation}\label{13}
U_{I}(t)\simeq1-\frac{\imath}{h}\int_{0}^{t}dt_{1}H_{ME}(t_{1}) -\frac{1}{h^{2}}\int_{0}^{t}dt_{2}\int_{0}^{t_{2}}dt_{1}H_{ME}(t_{2})H_{ME}(t_{1})
\end{equation}
First, we calculate
\begin{equation}\label{14}
U_{I}(t)|vac\rangle\simeq U_{vac}(t)|vac\rangle+U_{\alpha}(t)|\alpha\rangle+U_{\alpha\beta}(t)|\alpha\beta\rangle
\end{equation}
where
\begin{align}\label{15}
   U_{vac}(t)&=1-\frac{\imath t}{2h}\sum_{\alpha}\omega_{\alpha}^{2}\gamma_{\alpha}^{2}-\frac{1}{2h}\sum_{\alpha}\omega_{\alpha}^{3}\gamma_{\alpha}^{2}
   \int_{0}^{t}dt_{2}\int_{0}^{t_{2}}dt_{1}e^{-\imath\omega_{\alpha}(t_{2}-t_{1})}\sigma_{z}(t_{2})\sigma_{z}(t_{1})\nonumber \\
   U_{\alpha}(t)&=\frac{\imath}{\sqrt{2h}}\omega_{\alpha}^{3/2}\gamma_{\alpha}\int_{0}^{t}dt_{1}e^{\imath\omega_{\alpha}t_{1}}\sigma_{z}(t_{1}) \nonumber \\
   U_{\alpha\beta}(t)&=-\frac{1}{2h}\int_{0}^{t}dt_{2}\int_{0}^{t_{2}}dt_{1}\omega_{\alpha}^{3/2}\omega_{\beta}^{3/2}\gamma_{\alpha}\gamma_{\beta}
   e^{\imath(\omega_{\alpha}t_{1}+\omega_{\beta}t_{2})}\sigma_{z}(t_{2})\sigma_{z}(t_{1})
\end{align}
to find
\begin{equation}\label{16}
|\psi(t)\rangle=\sum_{n=1}^{2}e^{-\imath E_{n}t/h}|n\rangle\Big\{|vac\rangle\langle n|U_{vac}(t)|\psi\rangle+\sum_{\alpha}e^{-\imath\omega_{\alpha}t}|\alpha\rangle\langle n|U_{\alpha}(t)|\psi\rangle+\sum_{\alpha\beta}e^{-\imath(\omega_{\alpha}+\omega_{\beta})t}|\alpha\beta\rangle\langle n|U_{\alpha\beta}(t)|\psi\rangle\Big\}
\end{equation}
where we defined $\sigma_{z}(t)=e^{\imath H_{M}t/h}\sigma_{z}e^{-\imath H_{M}t/h}$. The contribution of the third term of (\ref{16}) in the probability of tunneling, being of order $\gamma^{4}$, and hence beyond the second order, can be securely dropped. The problem is thus reduced to the evaluation of matrix elements of $U_{vac}(t)$ and $U_{\alpha}(t)$. The diagonal matrix elements of $U_{vac}(t)$ are evaluated as
\begin{equation}\label{17}
\langle n|U_{vac}(t)|n\rangle=1-\frac{\imath t}{\hbar}\delta E_{n}^{(1)}+\frac{\imath}{\pi h}\sum_{m}|\sigma_{mn}|^{2}\int_{0}^{\infty}d\omega~J(\omega)\int_{0}^{t}dt'~\frac{1-e^{-\imath(\omega+\Omega_{mn})t'}}{\omega+\Omega_{mn}}
\end{equation}
where $J(\omega)$ is the spectral density of the environment, corresponding to a continuous spectrum of environmental frequencies, $\omega$, defined as
\begin{equation}\label{18}
J(\omega)=\frac{\pi}{2}\sum_{\alpha}\gamma_{\alpha}^{2}\omega_{\alpha}^{3}\delta(\omega-\omega_{\alpha})
\end{equation}
In the third term of (\ref{17}), the result of the integration remains unchanged if the integral is regarded as the principal-value integral which allows us to perform the $t'$-integration term by term as
\begin{align}\label{19}
\langle n|U_{vac}(t)|n\rangle=1-\frac{\imath t}{h}\Bigg\{\delta E_{n}^{(1)}-\frac{1}{\pi}
\sum_{m}|\sigma_{mn}|^{2}\wp\!\int_{0}^{\infty}\!d\omega\frac{J(\omega)}{\omega+\Omega_{mn}}\Bigg\}-\frac{1}{\pi h}\sum_{m}|\sigma_{mn}|^{2}\wp\!\int_{0}^{\infty}\!d\omega J(\omega)\frac{1-e^{-(\omega+\Omega_{mn})t}}{(\omega+\Omega_{mn})^{2}}
\end{align}
where the symbol $\wp$ denotes the principal-value integral. The quantity embraced by the braces in the second term on the right hand side coincides with $\delta E_{n}$ in (\ref{9}). So, the following expression is valid up to the second order
\begin{align}\label{20}
\langle n|U_{vac}(t)|n\rangle&\simeq e^{-\frac{\imath t}{\pi h}\delta E_{n}}\Bigg\{1-\frac{1}{\pi h}\sum_{m}|\sigma_{mn}|^{2}~\wp\int_{0}^{\infty}d\omega J(\omega)\frac{1-e^{-(\omega+\Omega_{mn})t}}{(\omega+\Omega_{mn})^{2}}\Bigg\}  \nonumber\\
  &=e^{-\frac{\imath t}{\pi h}\delta E_{n}}\Bigg\{1-\frac{1}{\pi h}\sum_{m}|\sigma_{mn}|^{2}\Bigg [\wp\int_{0}^{\infty}d\omega J(\omega)\Big[2\Big(\frac{\sin\{(\omega+\Omega_{rn})t/2\}}{\omega+\Omega_{mn}}\Big)^{2}-\imath\frac{\sin\{(\omega+\Omega_{mn})t\}}
  {(\omega+\Omega_{mn})^{2}}\Big]\Bigg]\Bigg\}
\end{align}
We assume that environmental cut-off frequency $\Lambda$ is much higher than the characteristic frequency of the molecule $\Omega$, so at times much higher than $\Omega^{-1}$, we approximate the first term of the integral on the second line by a delta function $\delta(\omega+\Omega_{mn})$. The result of the corresponding integral would be $J(\omega+\Omega_{mn})$, which is zero if $\Omega_{mn}\geq0$. Moreover, in the limit $\Delta\gg\delta$, we can safely approximate $\sigma_{nn}\sim0$. The diagonal elements of $U_{vac}(t)$ are then reduced to
\begin{equation}\label{21}
\langle 1|U_{vac}(t)|1\rangle\simeq e^{-\imath t\delta E_{1}/h}\Bigg\{1-\frac{\imath\sin^{2}(2\theta)}{\pi h}\wp\int_{0}^{\infty}d\omega J(\omega)\frac{\sin\{(\omega-\Omega_{12})t\})}{(\omega-\Omega_{12})^{2}}\Bigg\}
\end{equation}
and
\begin{equation}\label{22}
\langle 2|U_{vac}(t)|2\rangle\simeq e^{-\imath t\delta E_{2}/h}\Bigg\{1-\frac{1}{2}\Gamma_{2}t-\frac{\imath\sin^{2}(2\theta)}{\pi h}\wp\int_{0}^{\infty}d\omega J(\omega)\frac{\sin\{(\omega+\Omega_{12})t\})}{(\omega+\Omega_{12})^{2}}\Bigg\}
\end{equation}
where $\Omega_{12}=-(\Delta^{2}+\delta^{2})^{1/2}$.
The off-diagonal elements of $U_{vac}$ can be approximated as zero in the limit $\Delta\gg\delta$.\\
\indent Matrix elements of $U_{\alpha}(t)$ are obtained by
\begin{equation}\label{23}
\langle m|U_{\alpha}(t)|n\rangle=\frac{\imath}{\sqrt{2h}}\omega_{\alpha}^{3/2}\gamma_{\alpha}\sigma_{mn}
\frac{\sin{\{(\omega_{\alpha}+\Omega_{mn})t/2\}}}{(\omega_{\alpha}+\Omega_{mn})/2}e^{\imath(\omega_{\alpha}+\Omega_{mn})t/2}
\end{equation}
To obtain the explicit forms of matrix elements of (\ref{23}), one can follow a procedure pretty much the same as above. Especially, one can show that the diagonal elements of $U_{\alpha}(t)$ can be estimated as zero in the limit $\Delta\gg\delta$.\\
\indent We suppose that the initial state of the molecule is the left-handed state $|L\rangle$. The evolved state function of the whole system in the chiral basis of the molecule can be written as
\begin{equation}\label{24}
|\Psi(t)\rangle=\Big\{\sin{\theta}|\chi_{1}(t)\rangle-\cos{\theta}|\chi_{2}(t)\rangle\Big\}|L\rangle+
\Big\{\cos{\theta}|\chi_{1}(t)\rangle+\sin{\theta}|\chi_{2}(t)\rangle\Big\}|R\rangle
\end{equation}
with
\begin{equation}\label{25}
|\chi_{i}(t)\rangle=e^{-\imath E_{i}t/h}\Big\{|vac\rangle\langle i|U_{vac}(t)|L\rangle+\sum_{\alpha}e^{-\imath\omega_{\alpha}t}|\alpha\rangle\langle i|U_{\alpha}(t)|L\rangle\Big\}
\end{equation}
for $i=1,2$. We are interested in the probability of finding the molecule in the right-handed state, i.e.,
\begin{equation}\label{26}
P_{R}=|\langle R|\Psi(t)\rangle|^{2}=\cos^{2}{\theta}~\langle\chi_{1}(t)|\chi_{1}(t)\rangle+\sin^{2}{\theta}~\langle\chi_{2}(t)|\chi_{2}(t)\rangle
+\sin{(2\theta)}~\Re(\langle\chi_{1}(t)|\chi_{2}(t)\rangle)
\end{equation}
\section{IV. Results}
\begin{figure}[H]
  \subfigure[]{\includegraphics[scale=0.65]{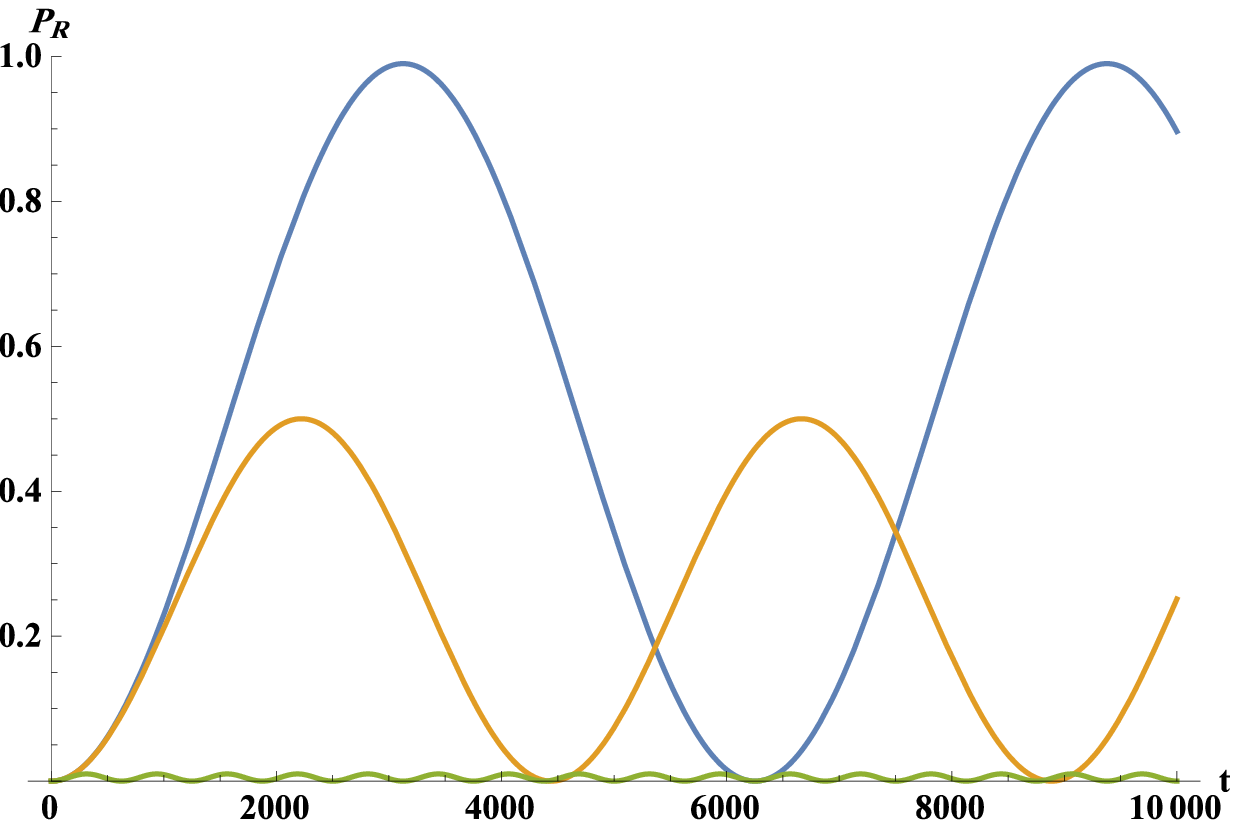}}\centering
  ~~~~~~
  \subfigure[]{\includegraphics[scale=0.75]{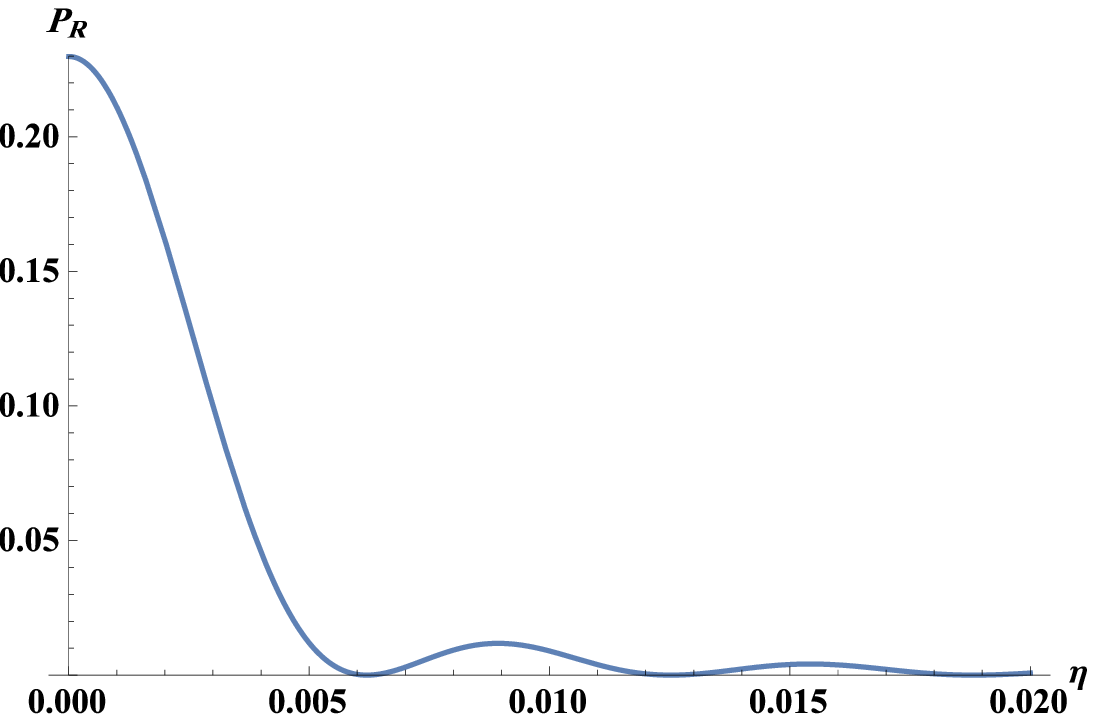}}\centering
  \caption{ The isolated chiral molecule: a) dynamics of the right-handed state for $\eta=10^{-4}$ (\textit{blue}), $\eta=10^{-3}$ (\textit{brown}), $\eta=10^{-2}$ (\textit{green}) b) probability of the right-handed state versus the asymmetry parameter $\eta$ at $t=1000$.}
  \label{Fig1}
\end{figure}

To examine the effective dynamics of the open chiral molecule, we first estimate the parameters relevant to our analysis. Two extreme cases can be distinguished: chiral molecules with very long tunneling times and chiral molecules with rapid tunnellings. For former, we have $\delta\gg\Delta$, giving $\theta\rightarrow0$, chiral states become stationary states and Hund's Paradox is resolved~\cite{Bar,HS3}. So, here we just focus on latter. The free parameters of the molecule can be chosen to be the effective mass $M$, the characteristic energy $U_{\mbox{\tiny${\mbox{\tiny$\circ$}}$}}$ and the characteristic length $R_{\mbox{\tiny${\mbox{\tiny$\circ$}}$}}$. The effective mass of a typical small chiral molecule (e.g., isomeric ammonia, NHDT) is estimated as $M\sim10^{-27}Kg$. The two-level approximation requires that $E_{th}\ll\Omega\tau_{\mbox{\tiny${\mbox{\tiny$\circ$}}$}}h\ll 1$, so considering the thermal energy at the room temperature as $k_{\mbox{\tiny$B$}}T\sim10^{-21}J$, we estimate the typical value of characteristic energy as $U_{\mbox{\tiny${\mbox{\tiny$\circ$}}$}}\sim10^{-19}J$. We estimate the characteristic length for a typical molecule close to molecular lengths as $R_{\mbox{\tiny${\mbox{\tiny$\circ$}}$}}\sim10^{-10}m$. The rest of the molecular parameters are estimated accordingly. We summarize relevant molecular parameters in TABLE I.
\begin{table}[H]
\begin{center}
\def\arraystretch{1.2}
\begin{tabular} {c c c c c c}
  Parameter &\hspace{0.4cm} Value &\hspace{0.4cm} Parameter &\hspace{0.4cm} Value &\hspace{0.4cm} Parameter &\hspace{0.4cm} value\\
  \hline
  $M$ &\hspace{0.4cm} $10^{-27}Kg$ &\hspace{0.4cm} $P_{\mbox{\tiny${\mbox{\tiny$\circ$}}$}}$ &\hspace{0.4cm} $10^{-23}Kg.m.s^{-1}$ &\hspace{0.4cm} $\Omega$ &\hspace{0.4cm} $10^{13}Hz$ \\
  $U_{\mbox{\tiny${\mbox{\tiny$\circ$}}$}}$ &\hspace{0.4cm} $10^{-19}J$ &\hspace{0.4cm} $\tau_{\mbox{\tiny${\mbox{\tiny$\circ$}}$}}$ &\hspace{0.4cm} $10^{-14}s$ &\hspace{0.4cm} $\Delta$ &\hspace{0.4cm} $10^{-3}$ \\
  $R_{\mbox{\tiny${\mbox{\tiny$\circ$}}$}}$ &\hspace{0.4cm} $10^{-10}m$ &\hspace{0.4cm} $h$ &\hspace{0.4cm} $0.1$ &\hspace{0.4cm} $\delta$ &\hspace{0.4cm} $\eta$ \\
\end{tabular}
\caption{Molecular parameters relevant for the model.}
\end{center}
\end{table}
\noindent Note that the value of tunneling frequency $\Delta$ coincides with the tunneling frequency of NHDT. Of particular interest is the value of reduced Planck constant obtained as $h\sim0.1$. The parameter $h$ quantifies the extent to which the isolated molecule exhibit quantum coherence. The situation in which $h\ll1$ is called the {\it quasi-classical} situation~\cite{Tak}.\\
\indent For an isolated chiral molecule, the probability of the right-handed state, according to (\ref{6}), as previously reported by Harris and Stodolsky~\cite{Har1,Har2} and recently by Bargue\~{n}o and co-workers~\cite{Lan1}, is implicitly independent of reduced Planck constant $h$. At the tunneling-dominant limit, corresponding to the symmetric double-well potential, the dynamics (blue plot in FIG.~\ref{Fig1},a) shows symmetric oscillations between two chiral states.
\noindent Introducing the asymmetry into the unitary dynamics reduces the symmetry of the oscillations, eventually confining them to the initial chiral state (brown and green plots in FIG.~\ref{Fig1},a). This oscillatory dynamics is the quantum signature of the system. At a definite time, the probability of the right-handed state is decreased quickly with the asymmetry parameter, $\eta$, (FIG.~\ref{Fig1},b). This is because the chiral interactions overcome tunneling process, and thus confine the molecule in the initial chiral state.\\
\indent A molecule, especially in a biological environment, is not really isolated. The environmental effects on the molecule may be classified into {\it decoherence} and {\it dissipation}. The former is the suppression of the relative phases (coherences) between the states of the molecule. The decoherence in the energy basis of the molecule is usually called dephasing. In our approach, the decoherence effect is characterized by the third term of (\ref{26}). The latter is the energy exchange with the environment leading to thermalization which is usually accompanied by decoherence. In our approach, first and second terms of (\ref{26}) are responsible for dissipation effect. In order to examine the dynamics of the open chiral molecule, first we should specify the type of the environment.
\subsection{Dilute Environment}
The interaction of the molecule with a dilute environment can be approximately accounted by two-body collisions. A gas phase is a generic example of such an environment. A gas phase is not actually a collection of harmonic oscillators. In fact, the fluctuating force produced by the gas particles onto molecule is not Gaussian (as it would be for oscillators), but rather looks like a series of random delta peaks. Nevertheless, if every collision is sufficiently weak, then fluctuating force could still approximately considered as Gaussian (when averaged over longer time scales, including many collisions). For the gas phase we obtain a simple expression for the right-handed state dynamics.\\
\indent For the coupling between the molecule and gas particles is weak, we have $\Gamma_{2}\ll\Omega$. At the temporal domain $\Omega^{-1}\ll t\ll\Gamma_{2}^{-1}$, we obtain
\begin{align}\label{27}
P_{1}(t)=\langle\chi_{1}(t)|\chi_{1}(t)\rangle&\simeq\sin^{2}(\theta)+\frac{\cos^{2}(\theta)\sin^{2}(2\theta)}{\pi h}\wp\int_{0}^{\infty}d\omega J(\omega)\frac{\sin^{2}\{(\omega+\Omega_{12})t\})}{(\omega+\Omega_{12})^{2}}\nonumber\\
&=\sin^{2}(\theta)+\cos^{2}(\theta)\Gamma_{2}t\simeq\sin^{2}(\theta)+\cos^{2}(\theta)(1-e^{-\Gamma_{2}t})
\end{align}
and
\begin{align}\label{28}
P_{2}(t)=\langle\chi_{2}(t)|\chi_{2}(t)\rangle&\simeq\cos^{2}(\theta)e^{-\Gamma_{2}t}+\frac{\sin^{2}(\theta)\sin^{2}(2\theta)}{\pi h}\wp\int_{0}^{\infty}d\omega J(\omega)\frac{\sin^{2}\{(\omega-\Omega_{12})t\})}{(\omega-\Omega_{12})^{2}}\nonumber\\
&\simeq\cos^{2}(\theta)e^{-\Gamma_{2}t}
\end{align}
One can easily show that $\cos^{2}(\theta)P_{1}(t)+\sin^{2}(\theta)P_{2}(t)\sim1/2$ and $P_{1}(t)P_{2}(t)\sim1$. So, the effect of the environment manifests itself through the dephasing factor $\langle\chi_{1}(t)|\chi_{2}(t)\rangle$ rather than the dissipation factors $P_{i}(t)$ ($i=1,2$). This is in accordance with the corresponding master equation in the dilute-phase limit of collisional decoherence~\cite{Gon,Lin,Horn}. The dephasing factor is obtained as
\begin{align}\label{29}
\langle\chi_{1}(t)|\chi_{2}(t)\rangle\simeq\frac{\sin(2\theta)}{2}\Bigg\{&e^{-\Gamma_{2}t/2}\exp\Bigg[\imath\Omega_{21}t+\imath(\delta E_{2}-\delta E_{1})t\nonumber\\&+\frac{\imath\sin^{2}(2\theta)}{\pi h}\int_{0}^{\infty}d\omega~J(\omega)\Big(\frac{\sin\{(\omega-\Omega_{12})t\}}{\omega-\Omega_{12}}
  -\frac{\sin\{(\omega+\Omega_{12})t\}}{\omega+\Omega_{12}}\Big)\Bigg]\nonumber\\&+\frac{4\imath\sin^{2}(2\theta)}{\pi h}\int_{0}^{\infty}d\omega J(\omega)\frac{\sin\{(\omega-\Omega_{12})t\}\sin\{(\omega+\Omega_{12})t\}}{\omega^{2}-\Omega^{2}_{12}}\Bigg\}
\end{align}
Finally, the right-handed probability in the temporal domain $\Omega^{-1}\ll t\ll\Gamma_{2}^{-1}$ is reduces to
\begin{align}\label{30}
P_{R}\sim\cos^{2}(\theta)P_{1}+\sin^{2}(\theta)P_{2}-\frac{\sin^{2}(2\theta)}{2}e^{-\Gamma_{2}t/2}\cos\Big\{(\Omega_{21}+\delta E_{2}-\delta E_{1})t+\zeta\Big\}
\end{align}
where
\begin{equation}\label{31}
\zeta=\frac{\sin^{2}(2\theta)}{h}\Big(J'(\Omega_{21})-\frac{J(\Omega_{21})}{\Omega_{21}}\Big)
\end{equation}
where we defined $J'(\omega)=dJ(\omega)/d\omega$. The dephasing effects of the gas phase are manifested in $e^{-\Gamma_{2}t/2}$ and $\xi$.\\
\indent In order to examine the dynamics explicitly, we should specify the spectral density of the gas phase. As Harris and Stodolsky demonstrated, the relevant dynamics of chiral states samples the velocity distribution of the gas molecules, which is strongly temperature-dependent~\cite{Har1,Har2}. Starting from a microscopic model of the collisions with the gas particles, we derive the spectral density
\begin{equation}\label{32}
J(\omega)=\int_{-\infty}^{\infty}dt~\langle X(t)X(0)\rangle_{E}e^{\imath\omega t}
\end{equation}
where $X(t)$ is the interaction-picture position operator of the environmental particles. Note that for simplicity we drop all normalization factors expressed by $2\pi$. The interaction process envisaged here is a sequence of collisions between gas particles and a heavier chiral molecule. We also assume that the collision doesn't lead to any internal transition of the molecule. The position operator of the environmental particles can be expanded as~\cite{Dob}
\begin{equation}\label{33}
X=\int d^{3}r~a(r)\bar{\rho}_{E}(r)
\end{equation}
where $a(r)$ is the interaction function and $\bar{\rho}_{E}(r)$ is the difference between the gas density operator $\rho_{E}(r)$ and its time-averaged value, assumed to be the uniform gas density $\rho$. If we substitute (\ref{33}) in (\ref{32}), we have
\begin{equation}\label{34}
J(\omega)=\int d^{3}q~|\tilde{a}(q)|^{2}S(q,\omega)
\end{equation}
where $\tilde{a}(q)$ is the Fourier transform of the interaction function
\begin{equation}\label{35}
\tilde{a}(q)=\int d^{3}r~a(r)e^{-\imath q.r}
\end{equation}
and we defined
\begin{equation}\label{36}
S(q,\omega)=\int dt\int d^{3}r~e^{\imath(\omega t-k.r)}\langle\bar{\rho}_{E}(0,0)\bar{\rho}_{E}(r,t)\rangle
\end{equation}
which is essentially the spectral density appeared in the first order theory of scattering~\cite{Tay}. The density operator of a gas of free particles is conveniently defined as
\begin{equation}\label{37}
\rho_{E}(r)=\frac{1}{V}\sum_{kq}e^{\imath q.r}b_{k+q}^{\dagger}b_{k}
\end{equation}
where $V$ is the normalization volume and $b_{k}$ is the annihilation operator. If we insert (\ref{37}) into (\ref{36}), we obtain
\begin{equation}\label{38}
S(q,\omega)=\rho\int d^{3}p~\bar{n}_{p}(1\pm\bar{n}_{p-hk})\delta(\omega-\frac{E_{p}-E_{hk}}{h})
\end{equation}
Here, $\rho$ is the density of the gas particles per unit volume, $\bar{n}_{hk}=\langle b_{k}^{\dagger}b_{k}\rangle$ is the Bose or Fermi distribution and $E_{p}=p^{2}/2$ is the single-particle energy. We are interested in the classical, non-degenerate regime, i.e,  $\bar{n}_{p}\ll1$, which leads to $\bar{n}_{p}(1\pm\bar{n}_{p-hk})\approx e^{-p^{2}/2E_{th}}$. After some mathematical manipulation, the spectral density becomes
\begin{equation}\label{39}
J(\omega)=\frac{\rho t_{c}}{R}e^{\omega t_{Q}}\int_{0}^{\infty}\frac{|\tilde{a}(q)|^{2}}{q}
e^{\scalebox{1}{$-(\frac{\omega^{2}t_{c}^{2}}{q^{2}}+\frac{t_{Q}^{2}q^{2}}{t_{c}^{2}})$}}
\end{equation}
where $R$ is the range of intermolecular interaction$, t_{c}=(R^{2}/2E_{th})^{1/2}$ and $t_{Q}=h/E_{th}$ are the classical and quantum correlation times, respectively. At room temperature, $t_{Q}\approx10^{-1}$, and for an interaction range $R>1$ and gas particles heavier than Hydrogen molecules, we estimate $t_{c}>1$. Thus, we ignore the term $t_{Q}^{2}/t_{c}^{2}$ ($<10^{-2}$). For a gaussian interaction function $|\tilde{a}(q)|^{2}=e^{-q^{2}}$, and assuming $t_{c}\gg t_{Q}$, the explicit form of the spectral density would be
\begin{equation}\label{40}
J(\omega)=J_{\mbox{\tiny${\mbox{\tiny$\circ$}}$}}\omega^{\frac{1}{2}} e^{-\frac{\omega}{\Lambda}}
\end{equation}
with coupling strength $J_{\mbox{\tiny${\mbox{\tiny$\circ$}}$}}\propto\rho E_{th}^{-3/4}$ and cut-off frequency $\Lambda=2(4t_{c}-t_{Q})^{-1}$. The spectral density (\ref{40}) shows a sub-ohmic frequency dependence with a temperature-dependent coupling strength and cut-off frequency. At the room temperature, assuming the dilute gas limit, we estimate $J_{\mbox{\tiny${\mbox{\tiny$\circ$}}$}}<1$, and $\Lambda>0.5$.\\
\subsection{Condensed Environment}
\begin{figure}[H]
  \subfigure[]{\includegraphics[scale=0.6]{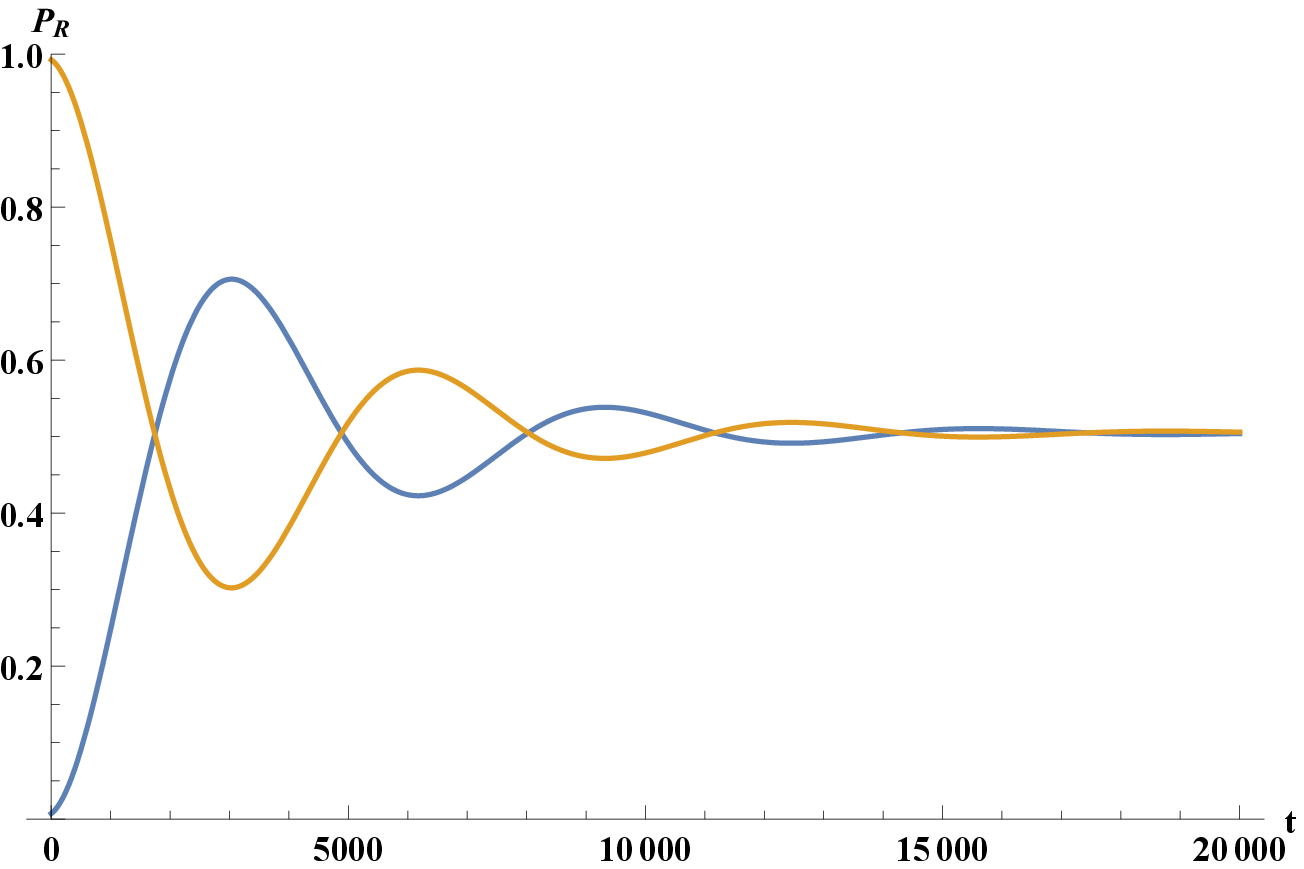}}\centering
  ~~~~
  \subfigure[]{\includegraphics[scale=0.6]{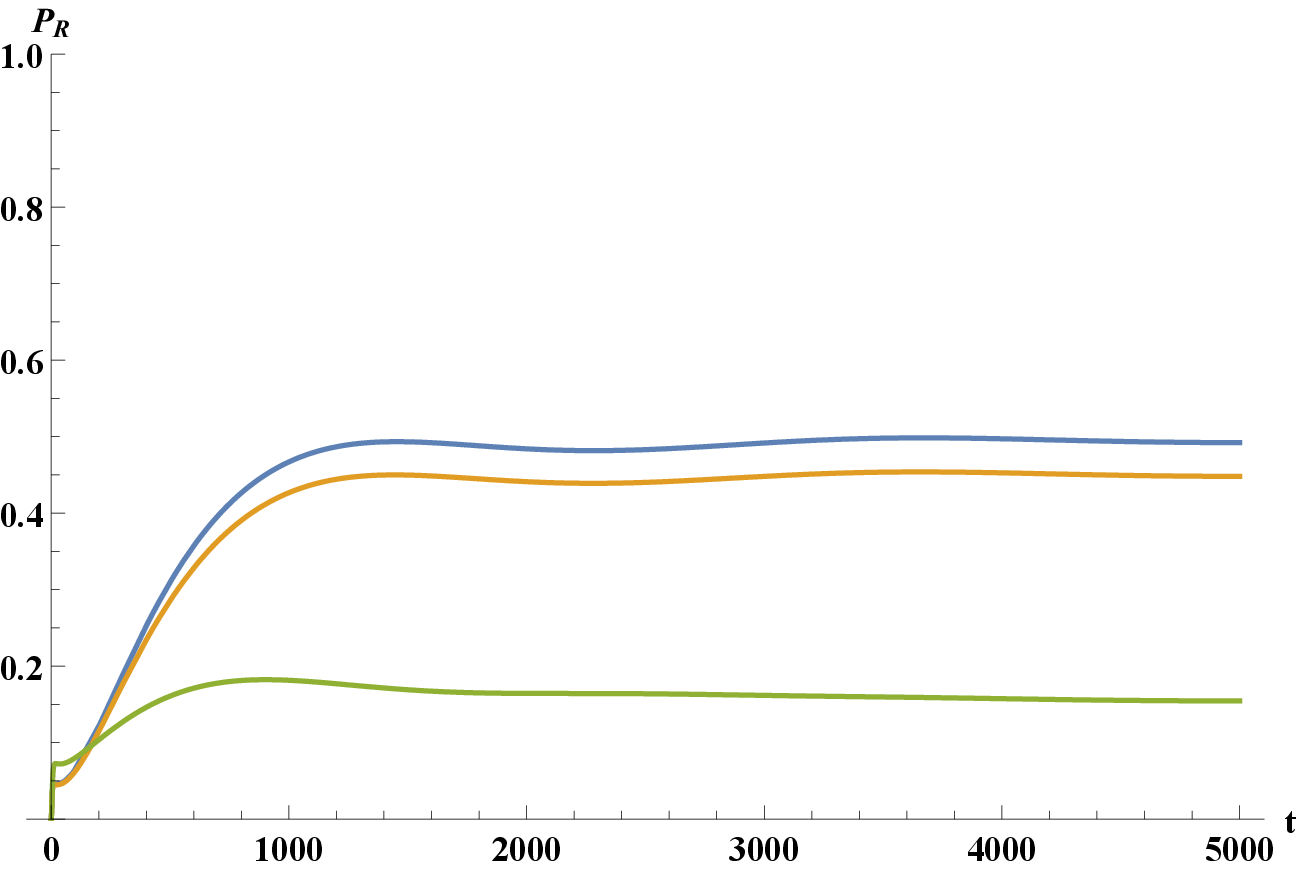}}\centering
  \caption{ The dynamics of the right-handed state of the open chiral molecule a) in the dilute phase at $|\delta|=10^{-5}$ for $\delta>0$ (blue) and $\delta<0$ (orange) b) in the condensed phase for $\delta=\pm10^{-5}$ (blue), $\pm10^{-4}$ (orange) and $\pm10^{-3}$ (green).}
  \label{Fig2}
\end{figure}

\begin{figure}[H]
  \subfigure[]{\includegraphics[scale=0.65]{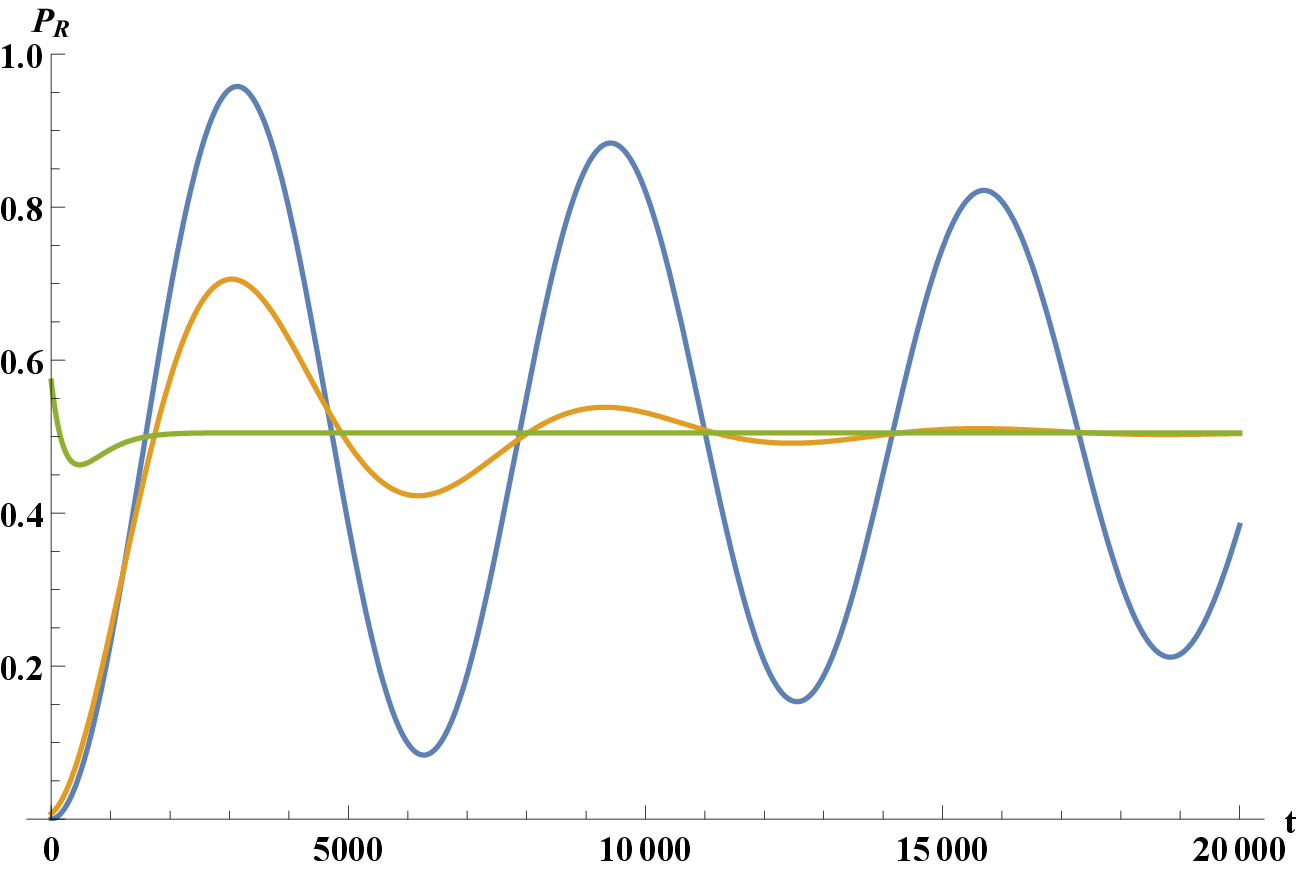}}\centering
  ~~~~
  \subfigure[]{\includegraphics[scale=0.65]{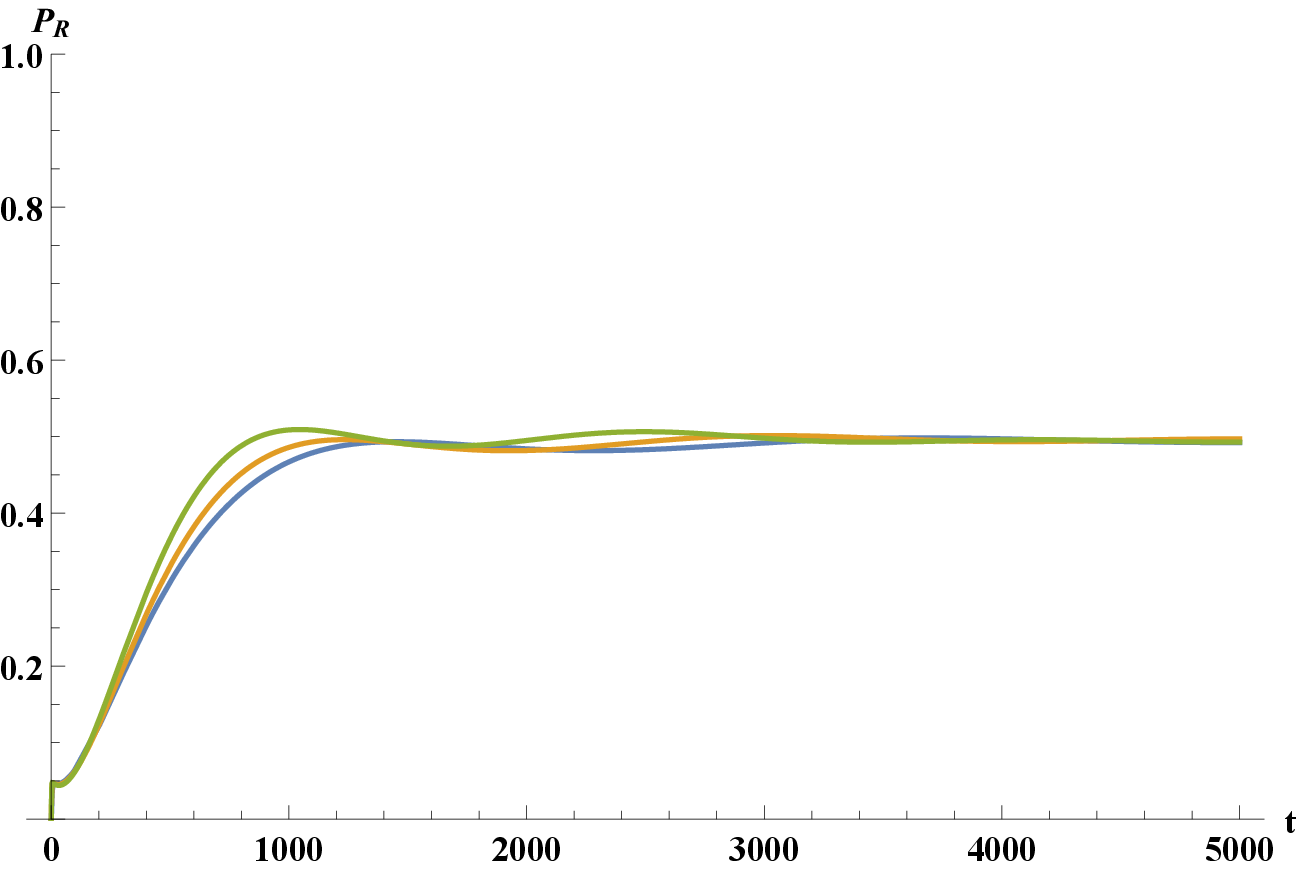}}\centering
  \caption{ The dynamics of the right-handed state of the open chiral molecule at $|\delta|=10^{-5}$ a) in the dilute phase for $J_{\mbox{\tiny${\mbox{\tiny$\circ$}}$}}=10^{-4}$ (blue), $10^{-3}$ (orange) and $10^{-2}$ (green) b) in the condensed phase for $J_{\mbox{\tiny${\mbox{\tiny$\circ$}}$}}=10$ (blue), $20$ (orange) and $30$ (green).}
  \label{Fig3}
\end{figure}

\begin{figure}[H]
  \subfigure[]{\includegraphics[scale=0.65]{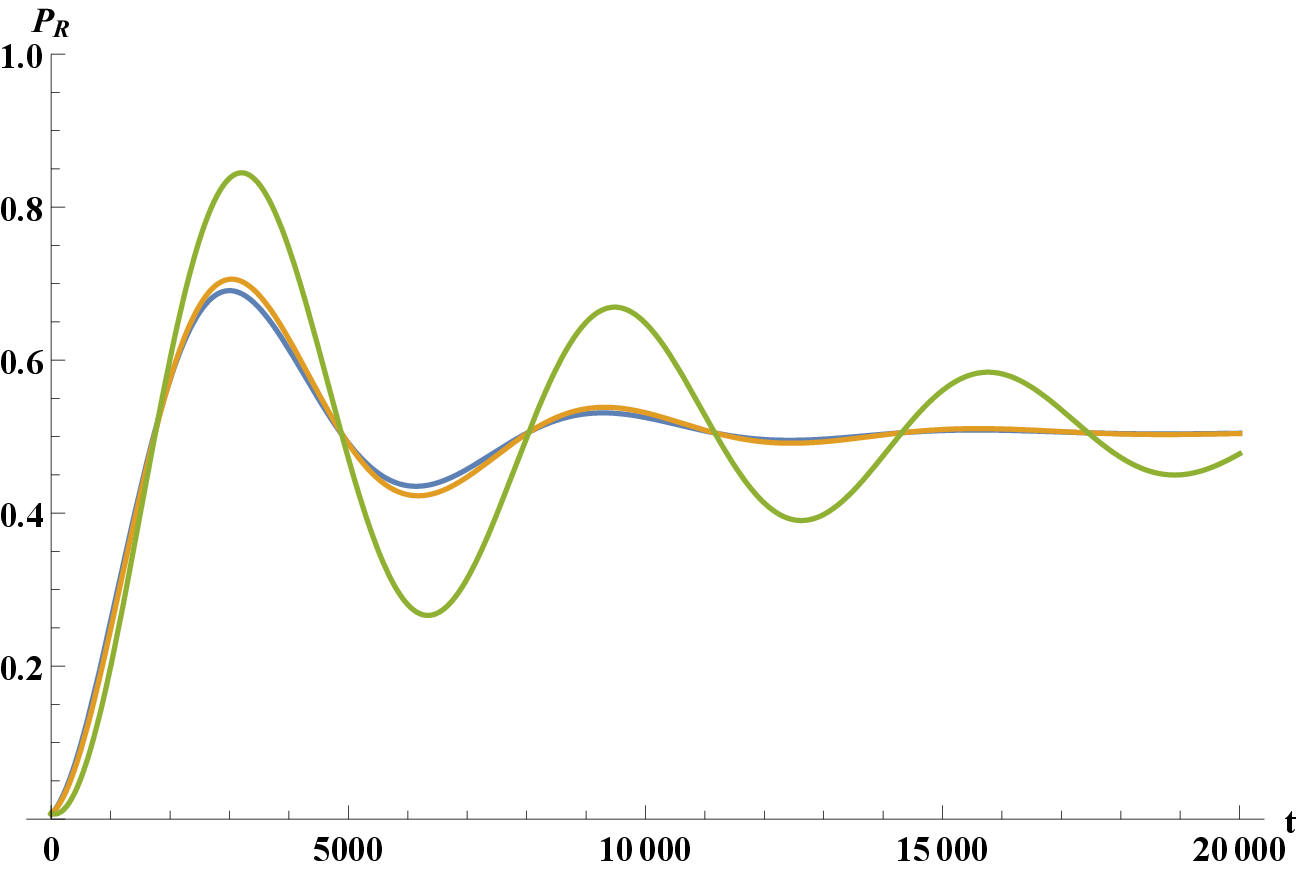}}\centering
  ~~~~
  \subfigure[]{\includegraphics[scale=0.65]{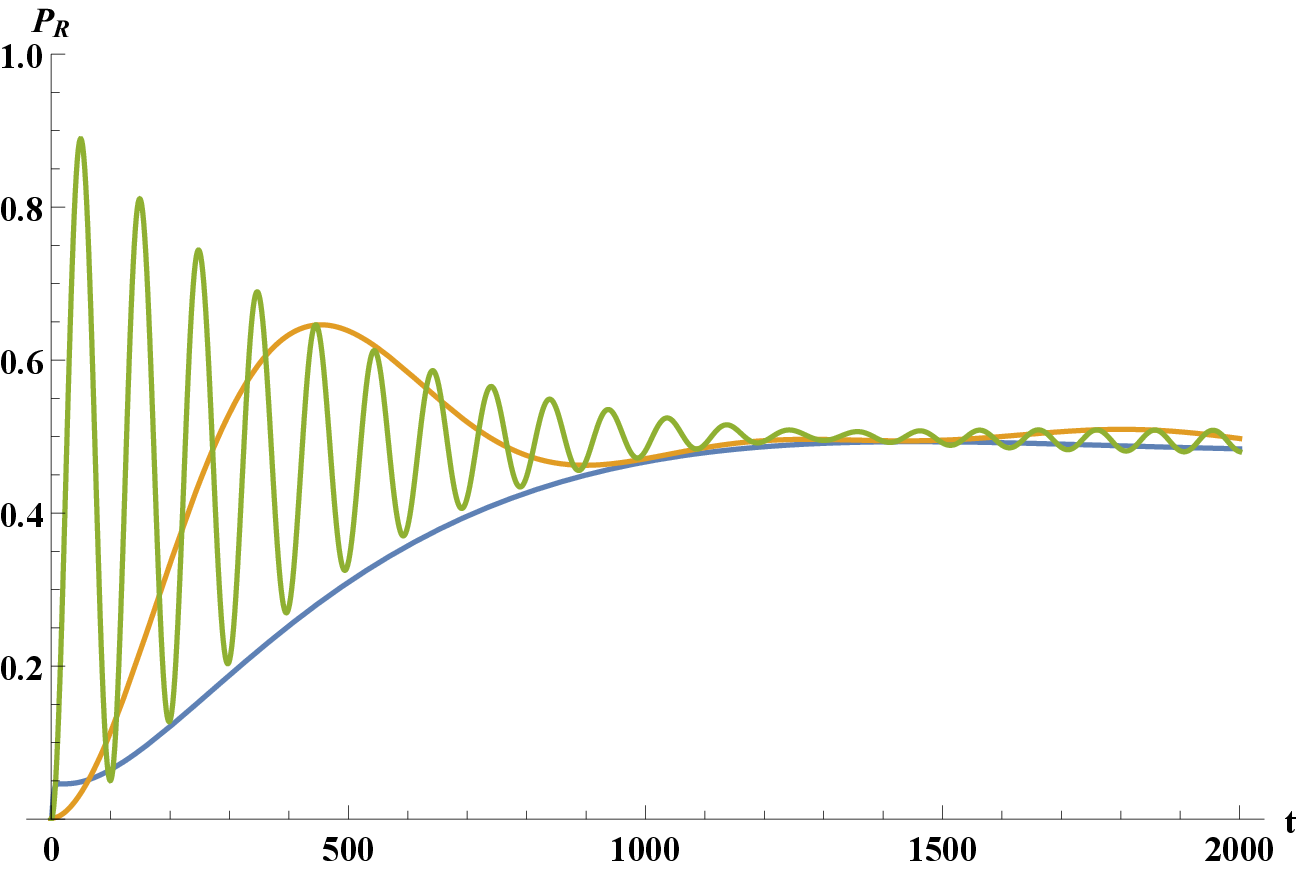}}\centering
  \caption{The dynamics of the right-handed state of the open chiral molecule at $|\delta|=10^{-5}$ a) in the dilute phase for $\Lambda=10^{-1}$ (blue), $10^{-2}$ (orange) and $10^{-3}$ (green) b) in the condensed phase for $\Lambda=10^{-2}$ (blue), $10^{-1}$ (orange) and $1$ (green).}
  \label{Fig4}
\end{figure}

In a condensed environment, since the medium is always present, the idea of collision is inapplicable. The properties of such an environment can be encapsulated in an ohmic spectral density with an exponential cutoff $\Lambda$ as
\begin{equation}\label{41}
J(\omega)=J_{\mbox{\tiny${\mbox{\tiny$\circ$}}$}}\omega e^{-\frac{\omega}{\Lambda}}
\end{equation}
in which $J_{\mbox{\tiny${\mbox{\tiny$\circ$}}$}}$ is a measure of the system-environment coupling strength. A condensed environment is more appropriate to represent a biological environment. The interaction of the molecule with a biological environment is, by nature, complex. We assume that the chiral molecule is surrounded by a uniform polar solvent. For simplicity, we describe the solvation process by the well-known Onsager model~\cite{Ons}. The chiral molecule is treated as a point dipole which is surrounded by a spherical cage of polar solvent molecules with Onsager radius $a$, which is typically the size of the molecule. For a Debye solvent, the parameters of the spectral density can be written as~\cite{GiMc}
\begin{equation}\label{42}
J_{\mbox{\tiny${\mbox{\tiny$\circ$}}$}}=\frac{(\Delta\mu)^{2}}{4\pi\epsilon_{\mbox{\tiny${\mbox{\tiny$\circ$}}$}}a^{3}}
\frac{6(\epsilon_{s}-\epsilon_{\infty})}{(2\epsilon_{s}+1)(2\epsilon_{\infty}+1)\Lambda}
\end{equation}
and
\begin{equation}\label{43}
\Lambda=\frac{1}{\tau_{D}}\frac{2\epsilon_{s}+1}{2\epsilon_{\infty}+1}
\end{equation}
where $\Delta\mu$ is the difference between the dipole moment of the molecule in the ground and excited states, $\epsilon_{s}$ and $\epsilon_{\infty}$ are the static and high-frequency dielectric constants of the solvent, and $\tau_{D}$ is the Debye relaxation time of the solvent. For water as the most prevalent solvent we have $\epsilon_{s}=78.3$, $\epsilon_{\infty}=4.21$ and $\tau_{D}=8.2ps$~\cite{Afs}. Using these parameters, we obtain (dimension-less) cut-off frequency as $\Lambda\approx0.01$. Note that if $\Lambda\ll|\Omega_{12}|$, the environmental particles cannot resolve the molecular states and thus the quantum oscillations resulted from tunneling process preserves. Also, if we measure the cavity size and change in dipole moment in angstroms and Debye respectively, then we obtain coupling strength as $J_{\mbox{\tiny${\mbox{\tiny$\circ$}}$}}=22\frac{(\Delta\mu)^{2}}{a^{3}}$. For a typical small molecule with radius $\approx1{\AA}$, a dipole moment change of just $\sim0.2D$ is sufficient to make $J_{\mbox{\tiny${\mbox{\tiny$\circ$}}$}}>1$, and this condition seems likely to be met for most small molecules~\cite{Rei}. As an example, for $NHDT$ with $\Delta\mu\approx0.6D$~\cite{Jac}, we have $J_{\mbox{\tiny${\mbox{\tiny$\circ$}}$}}\approx10$.\\
\indent Now, we compare the dynamics in a dilute phase with that of a condensed phase at short- and long-time limits.\\
\indent {\bf Long-Time Dynamics} The equilibrium state is obviously reached in the dilute and condensed phases at different rates. The condensed-phase equilibrium, as expected, occurs faster than the dilute-phase one (FIG.~\ref{Fig2}). Of course, the value of the rates depends on the details of the corresponding spectral density. For NHDT molecule in hydrogen gas and water, the equilibrium is reached after $\sim10000$ $(10^{-11}s)$ and $\sim1000$ $(10^{-10}s)$, respectively. In the dilute phase, the chiral interactions are relatively weak and the chiral molecule in question has a high tunneling rate, so the dynamics is confined to the tunneling-dominant limit where $\delta\ll\Delta$. In the condensed phase, however, the short-range intermolecular chiral interactions would be significant. At the tunneling-dominant limit, in both dilute and condensed phases, the equilibrium state is a racemic mixture (FIG.~\ref{Fig2},a and blue curve of FIG.~\ref{Fig2},b). Introducing the asymmetry into the condensed-phase dynamics leads to the localization. In the condensed phase, the chiral interactions confine the molecule in the initial chiral state (green curve of FIG.~\ref{Fig2},b). These results are in agreement with those of master equation approach ( for dilute phase~\cite{Gon}, for condensed phase~\cite{Tir}), and Langevin approach~\cite{Lan1} for condensed phase. In both phases, the equilibrium state is independent of the chirality of the chiral interactions (FIG.~\ref{Fig2}). Also, the dilute- and condensed-phase dynamics in our work is compatible respectively with the weak- and strong-damping regimes of Spin-Boson approach~\cite{Sto2,Sto3}. Note that the equilibrium state, as Grifoni and co-workers predicted for the driven Spin-Boson model using the path integral method~\cite{Gri}, is independent of the initial state of the chiral molecule. Obviously, the rate at which the equilibrium state is reached in both phases depends on the coupling strength $J_{\mbox{\tiny${\mbox{\tiny$\circ$}}$}}$. In the dilute phase, as expected, the equilibrium rate is increased with the coupling strength. However, in the condensed phase, the equilibrium rate is approximately independent of the coupling strength (FIG.~\ref{Fig3}). This is because of the fact that the the number of collisions in the condensed phase are already reached its asymptotic value and thus increasing the coupling strength doesn't change the equilibrium rate. The equilibrium rate also depends on the cut-off frequency of the environment $\Lambda$. In both phases, the equilibrium rate is decreased with the cut-off frequency (FIG.~\ref{Fig4}).
{\bf The short-time dynamics} The short-time dynamics in the dilute phase is oscillatory and strongly dependent on the chirality of the chiral interactions (FIG.~\ref{Fig2},a). In the condensed phase, however, the quantum characteristics of the molecule are completely suppressed by to the strong interactions with the environment (FIG.~\ref{Fig2},b).
\section{V. Conclusion}
The dynamics of chiral states of chiral molecules is conveniently examined in the dilute and condensed phases, respectively, by scattering approach (collisional decoherence) and Born-Markov master equation. Two approaches being mathematically different, the extension of the molecular dynamics from the dilute phase to the condensed phase is not straightforward. Here, we examined the dynamics of chiral states, corresponding to the left- and right-handed states of an asymmetric double-well potential, in both phases by a unified approach using the time-dependent perturbation theory. Two phases are described as a harmonic environment. The spectral density of the dilute phase, exemplified by an inert background gas, is temperature-dependent and sub-ohmic (\ref{40}), while the spectral density of the condensed phase, manifested as water, is temperature-independent and ohmic (\ref{41}). As our analysis implies, the dynamics in the dilute-phase cannot be extended to the dynamics of the condensed phase by increasing the coupling strength. This suggest that in the condensed phase, the unexpected features may emerge in the dynamics. Especially, we showed that the chiral interactions in the condensed phase may induce localization in the sense of the quantum Zeno effect (FIG.\ref{Fig2},b), while the dilute-phase dynamics always leads to racemization (FIG.\ref{Fig2},a). Also, in the short-time limit, the dilute-phase dynamics strongly depends on the initial state of the molecule (FIG.\ref{Fig2},a) and on the coupling strength (FIG.\ref{Fig3},a), whereas the condensed-phase dynamics is not. Our results are compatible with the results of the relevant works in the collisional decoherence, the master equation and Langevin approaches.
\section{Acknowledgement}
F.T.G acknowledges the financial support of Iranian National Science Foundation (INSF) for this work.

\end{document}